\documentclass{pasj00}
\twocolumn
\begin{document}
\SetRunningHead{Author(s) in page-head}{Running Head}
\Received{yyyy/mm/dd}
\Accepted{yyyy/mm/dd}

\title{Detection of Iron Emission in the $z$ = 5.74 QSO SDSSp 
J104433.04-012502.2 \thanks{Based on data collected at Subaru Telescope, which is operated by the National Astronomical Observatory of Japan. }}

\author{Kentaro \textsc{Aoki} \thanks{Present address: Subaru Telescope, National Astronomical Observatory of Japan, 650 North A'ohoku Place, Hilo, Hawaii 96720 U.S.A.} }
\affil{Institute of Astronomy, School of Science, The University of Tokyo, Mitaka, Tokyo 181-0015}
\email{kaoki@ioa.s.u-tokyo.ac.jp}
\author{Takashi \textsc{Murayama}}
\affil{Astronomical Institute, Tohoku University, Aoba-ku, Sendai, Miyagi 980-8578}
\email{murayama@astr.tohoku.ac.jp}
\and
\author{Kiyomi {\sc Denda}}
\affil{National Astronomical Observatory of Japan, 2-21-1 Osawa, Mitaka, Tokyo 181-8588}

\KeyWords{galaxies: active --- galaxies: evolution ---  quasars: individual (SDSSp J104433.04-012502.2)}

\maketitle

\begin{abstract}
We obtained near-infrared spectroscopy of the $z$=5.74 QSO, SDSSp J104433.04-012502.2
with the Infrared Camera and Spectrograph of the Subaru telescope.
The redshift of 5.74 corresponds to a cosmological age of 1.0 Gyr for 
the current $\Lambda$-dominated cosmology.
We found a similar strength of the Fe \emissiontype{II} (3000-3500 \AA) emission 
lines in SDSSp J104433.04-012502.2 as in low redshift QSOs.
This is the highest redshift detection of iron.
We subtracted a power-law continuum from the spectrum and fitted model 
Fe \emissiontype{II} emission and Balmer continuum.
The rest equivalent width of
Fe \emissiontype{II} (3000-3500 \AA) is $\sim 30$ \AA~ which is similar to those of low redshift QSOs measured by the same manner.
The chemical enrichment models that assume the life time of the progenitor of 
SNe Ia is 
longer than 1 Gyr predict that weaker Fe \emissiontype{II} emission than low redshift.
However, none of the observed high redshift ($z > 3$) QSOs show a systematic decrease of Fe \emissiontype{II} emission compared with low redshift QSOs.
This may due to a shorter lifetime of SNe Ia in QSO nuclei than 
in the solar neighborhood.
Another reason of strong Fe \emissiontype{II} emission at $z=5.74$ may be longer cosmological age due to smaller $\Omega _{\rm M}$.
\end{abstract}

\section{Introduction}
The enrichment of $\alpha$-elements (e.g., O and Mg) in the universe has different time scales 
from that of iron-peak elements.
While both $\alpha$-elements and iron are produced by SNe II, SNe Ia produce mostly iron.
The progenitors of SNe Ia are intermediate mass stars in binary systems and 
have a lifetime of $\sim 1$ Gyr although there is still debate on the 
progenitors of SNe Ia and their lifetime 
(e.g. Branch 1998 and Hillebrandt and Niemeyer 2000 for recent reviews).
The 1 Gyr lifetime of the progenitors of SNe Ia is longer 
than those of SNe II, which are massive stars. 
The iron/alpha abundance ratio shows a rapid increase when SNe Ia start to 
explode.
The [O/Fe] - [Fe/H] relation in the solar neighborhood is reproduced by that 
scenario assuming a 0.5-3 Gyr lifetime of the progenitors of SNe Ia (Yoshii et al. 1996; Kobayashi et al. 1998).
Among high redshift objects, except for metal absorption-line systems, 
QSOs' spectra show both iron and $\alpha-$element spectral features.
In particular, both Mg \emissiontype{II} $\lambda2798$ \AA~and Fe \emissiontype {II} UV (2000-3000 \AA) and Fe \emissiontype{II} (3000-3500 \AA) emission-lines are
emitted from the same ionized zone of the broad line region gas, 
and their similar wavelengths mean that extinction is similar.
Therefore we can use the intensity ratio of their fluxes as an indicator of iron/alpha elements abundance ratios
(Wills et al. 1985; Hamann \& Ferland 1993).
Comparison of the observed intensity ratio, $I$(Fe \emissiontype{II})/$I$(Mg \emissiontype{II}), in 
low redshift QSOs with results of model calculations of the broad line region suggests iron 
overabundance by a factor of $\sim 3$ at low redshift (Wills et al. 1985).
The iron overabundance of low redshift QSOs is reproduced by a QSO chemical enrichment model assuming rapid star formation and a delay of SNe Ia events (Hamann \& Ferland 1993). \par
Recently, more than 20 high redshift ($3.3 < z < 4.7$) QSOs have been 
observed with near-infrared spectroscopy and 
Mg \emissiontype{II} and Fe \emissiontype{II} UV and Fe \emissiontype{II} (3000-3500 \AA) emission-lines were measured
(Kawara et al. 1996; Murayama et al. 1999; Thompson et al. 1999; Dietrich and Hamann 2001;
Dietrich et al. 2002).
Their $I$(Fe \emissiontype{II} (UV))/$I$(Mg \emissiontype{II}) are 7--10 and comparable to that of low redshift 
QSOs (Wills et al. 1985).
These ratios are interpreted as suggesting QSO ages $ > 1$ Gyr in terms of the 
chemical enrichment model (Hamann \& Ferland 1993; Yoshii et al. 1998).
The chemical enrichment model (Yoshii et al. 1998) indicates
that the $I$(Fe \emissiontype{II} (UV+optical))/$I$(Mg \emissiontype{II}) remains less 
than 2 until 1 Gyr has passed since first star formation occurred.
After 1 Gyr has passed, $I$(Fe \emissiontype{II} (UV+optical))/$I$(Mg \emissiontype{II}) 
rapidly increases to $\sim 12$ at 1.5 Gyr.
The highest redshift for which $I$(Fe \emissiontype{II})/$I$(Mg \emissiontype{II}) was 
observed was 4.7 (Thompson et al. 1999; Dietrich and Hamann 2001)
which corresponds to age of 1.3 Gyr in a cosmology with the current most likely cosmological parameters
\footnote{We adopt $\Omega_{\rm M}=0.3, \Omega_{\Lambda}=0.7, 
H_0=65$ km s$^{-1}$ Mpc$^{-1}$ throughout the paper 
(Schmidt et al. 1998; Perlmutter et al. 1999;de Bernardis et al. 2000;
Melchiorri et al. 2000).}.
The large value of the $I$(Fe \emissiontype{II})/$I$(Mg \emissiontype{II}) suggests that the first star formation occurred at $z \gtrsim 10$ which corresponds to 0.5 Gyr (Dietrich and Hamann 2001).  
\par
The QSO at $z$ = 5.74, SDSSp J104433.04-012502.2 
(hereafter SDSS 1044-0125; Fan et al. 2000) was discovered in 2000 spring by the Sloan Digital Sky Survey (York et al. 2000). 
It is a luminous QSO 
($L_{\nu, 2500}$$=5.2 \times 10^{31}$ erg s$^{-1}$ Hz$^{-1}$) and the highest 
redshift object as of February 2001 when the observation was made.
Maiolino et al (2001) have found a broad absorption line in its spectrum 
suggesting that heavy absorption by gas with a column density 
$N_{\rm H} > 10^{24}$ cm$^{-2}$ is the reason for its X-ray faintness (Brandt et al. 2001).
Goodrich et al. (2001) made moderate resolution near infrared spectroscopy and revised 
the redshift to 5.74 measuring the peak of C \emissiontype{IV} emission-line.
Since its redshift of 5.74 corresponds to a cosmological age of 1.0 Gyr,
SDSS 1044-0125 is less than 1 Gyr old and the progenitor of SNe Ia
in the QSO would not have exploded yet.
The Fe \emissiontype{II} emission of SDSS 1044-0125 is expected to be weak.
Mg \emissiontype{II} emission line is difficult to observe because it is redshifted
to a spectral region of low atmospheric transparency.

\section{Observations and Data Reduction}
Near-infrared ($J, H,$ and $K$ band) spectra of SDSS 1044-0125 were obtained 
with the Infrared Camera and Spectrograph (IRCS; Kobayashi et al. 2000) 
at the Cassegrain focus of the Subaru telescope on the nights of 2001 February 4 and 5 (UT).
Both nights were photometric, but the seeing was unstable.
It changed from \timeform{0.4''} to \timeform{1.2''} in $K$ band.
The projected pixel size of the $1024^{2}$ ALADDIN II array was \timeform{0.058''}
along the slit, and 3.7, 4.7 and 6.1 \AA~ along the dispersion direction in $J, H,$ and $K$ band, respectively. 
A slit width of \timeform{0.9''} was used resulting in resolutions in $J, H$, and 
$K$ band of 74, 117, and 198 \AA~FWHM, respectively.
We obtained 12, 12 and 64 exposures, each 300 sec integration, in $J, H$, and $K$ band, respectively, 
shifting the position of the QSO along the slit at intervals of \timeform{7''} between each integration.
Total integration times on the QSO were 3600, 3600, and 19200 sec in $J, H,$ and $K$ band respectively.
Observations of a nearby G5 star, HD 93019, were made before and after the observation of the QSO at approximately same airmass to correct the spectra for telluric absorption, and for flux calibration.
The spectra of a halogen lamp and a Argon lamp were obtained for flat field and wavelength calibration, respectively.
\par

Data were reduced using IRAF\footnote{Image Reduction and Analysis Facility (IRAF) is distributed by the National Optical Astronomy
Observatories, which are operated by the Association of Universities for Research in Astronomy, Inc.,
under cooperative agreement with the National Science Foundation. }.
Dark frames of the same integration times as for the QSO and 
the star, HD 93019, were subtracted from the data frames.
The following procedures were separately applied for each wavelength band.
After masking of bad pixels, flat fielding was done with an averaged and normalized flat frame.
Sky was removed by subtracting an adjacent exposure with the QSO in a different 
position along the slit.
The sky-subtracted QSO frames at the same position along the slit but different airmass were averaged because the low S/N of a single frame hampered a check on whether telluric absorption features were removed. 
Although S/N of the QSO's spectra were improved due to averaging, 
the continuum of the QSO in the averaged frame was not bright enough to trace its position.
Therefore we first traced the spectra of the telluric absorption correction star, HD 93019. 
This trace was used for extraction of the QSO's spectra and the comparison spectra.
The extracted one dimensional comparison spectra were used for wavelength calibration of the QSO's extracted one dimensional spectra. 
After wavelength calibration of the one dimensional spectra, the spectra of the QSO at different position along the slit were combined to produce the QSO's spectrum in each wavelength band.
The spectra of HD 93019 were extracted from the sky subtracted frames and wavelength calibrated using one dimensional extracted comparison spectra.
The spectra of HD 93019 obtained at different position along the slit but same airmass were combined to give a spectrum at each airmass in each wavelength band.
For relative flux calibration, we divided the spectra of HD 93019 by a 5560 K blackbody spectrum to make relative sensitivity functions including telluric absorption.
Since the QSO spectra obtained at different airmasses were combined into a single spectrum, we averaged relative sensitivity functions at different airmasses to correct telluric absorption and instrument sensitivity.
The S/N of the QSO spectra in $J$, $H$, and $K$ band are $\sim5$, $\sim2$, and $\sim12$, respectively.
\par
The detector covered between 1.18 and 1.38$\mu$m in $J$ band, however, the 
sensitivity of the grism used was low shortward of 1.24 $\mu$m and 
longward of 1.35 $\mu$m and the signal from the QSO in that wavelength range was almost nothing.
We therefore use the region between 1.24 $\mu$m and 1.35 $\mu$m for the $J$ band spectrum.

\section{Results}
Figure 1 shows the rest-frame spectra of SDSS 1044-0125 in $J$ and 
$K$ band together with the spectra obtained by Maiolino et al. (2001). 
\begin{figure}
  \begin{center}
    \FigureFile(70mm,20mm){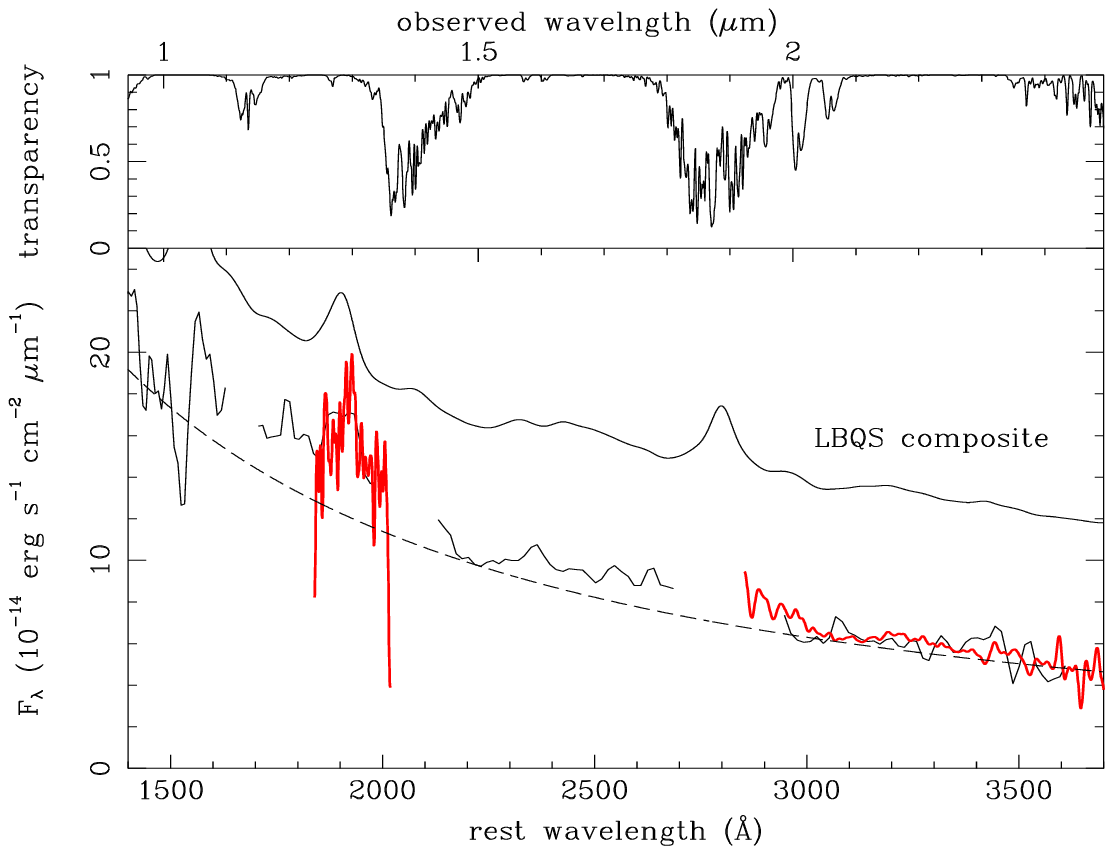}
  \end{center}
  \caption{(Upper panel) The atmospheric transmission curve. These data,
produced using the program IRTRANS4, were obtained  from the UKIRT www page. 
(Lower panel) Our spectra of SDSS 1044-0125, which is smoothed and shifted 
to the rest wavelength by the redshift of 5.74 (red line) and 
the reslut by Maiolino et al. (black line). 
The F$_{\lambda}$ scale of our SDSS 1044-0125 spectra is shifted to the spectra obtained
by Maiolino et al. (2001).
The smoothed LBQS composite spectrum is plotted for comparison.
It is shifted for clearity by 8.
The dashed line is a power-law continuum 
($F_{\nu} \propto \nu^{-\alpha}$) with $\alpha=$0.54. 
See text for details.} \label{fig1}
\end{figure}
The spectra of SDSS 1044-0125 were deredshifted by adopting $z$=5.74 (Goodrich et al. 2001) and were smoothed 4.8 \AA~ in $J$, 13.5 \AA~ in $K$ band, respectively. 
Since our spectra were obtained under unstable seeing conditions, 
light from the QSO falling into the \timeform{0.9''} slit varied among the
exposures.
The absolute flux density scale of the $J$ band to $K$ band spectrum is unknown.
Maiolino et al. (2001) simultaneously obtained $J$, $H$, and $K$ band spectra of SDSS 1044-0125.
We scaled our $J$ and $K$ band spectra to the spectra from Maiolino et al. (2001) at 1850-1965 \AA~ in $J$ band and 3000-3500 \AA~in $K$ band, respectively.
A emission feature is recognized at 1900 \AA~ in the rest frame.
This feature is a complex of Si \emissiontype {III}] and C \emissiontype{III}] as found by Maiolino et al. (2001).
Since our $H$ band spectrum of SDSS 1044-0125 is very low S/N, we do not use the $H$ band data.
As shown by Maiolino et al. (2001), the spectra of SDSS 1044-012 is similar to the Large Bright Quasar Survey (LBQS) composite spectrum \footnote{http://sundog.stsci.edu/first/QSOComposites/} using total data set 
(Brotherton et al. 2001), which is a low redshift optically selected QSO template.
The spectrum of SDSS 1044-0125 between 2900 \AA~ and 3400 \AA~ is similar to the LBQS composite spectrum.
The bump between 2900 \AA~ and 3000 \AA~ is Fe \emissiontype{II} emission (Vestergaard and Wilkes 2001) 
and increase of flux from 3000 \AA~ to 3400 \AA~ is also due to Fe \emissiontype{II} emission and the Balmer continuum (Wills et al. 1985; Verner et al. 1999).
\par
We measured Fe \emissiontype{II} emission-line strength as follows.
Since our spectrum covers limited wavelength range,
we determined the power-law continuum ($F_{\nu} \propto \nu^{-\alpha}$)
at $1465 $\AA~ and $2231 $\AA~in the spectrum of Maiolino et al. (2001).
After subtracting the power-law continuum from the $K$ band spectrum 
of SDSS 1044-0125, 
we assume that the SDSS1044-0125 spectrum can be fitted with a Fe \emissiontype{II} template and the Balmer continuum, so that
\[ F_{\lambda, SDSS}= a F_{\lambda, Fe II} + b F_{\lambda, Balmer~ continuum}. \]
We have done $\chi^2$ fitting between 2855 \AA, which is the shortest end of the $K$ band spectrum, and 3600 \AA, and derived the scaling parameter, $a$ and $b$.
The Fe \emissiontype{II} template was adopted from Wills et al. (1985).
It was calculated assuming the hydrogen density, $n_{H} = 10^{9.5}$ cm$^{-3}$, 
the ionization parameter, $U_{1Ryd} = 1.2 \times 10^{8}$ cm sec$^{-1}$.
The optical depth at 2343 \AA, $\tau$(2343) $\times$ the turbulent velocity, $V_{t}$ was 
asssumed to be constant.
We choosed the template under the condition of $\tau$(2343) = 5 $\times 10^4$ and
$V_{t} = 7$ km sec$^{-1}$ because the relative strength between 
Fe \emissiontype{II} multiplets were the most similar to the data of SDSS 1044-0125 among the templates available.
We broadened the Fe \emissiontype{II} template by convolving it with a Gaussian of FWHM=6000
km sec$^{-1}$ which is the line width of the C \emissiontype{IV} emission-line
(Goodrich et al. 2001).
The Balmer continuum assumed was a partially 
optical thick case given by Grandi (1982), with $\tau_{BE}=1.0$, $\nu_{BE}=3646$\AA~ and $T=10000$ K,
\[ F_{\nu} \propto B_{\nu}(T)(1-e^{-\tau})~ {\rm with}~ \tau=\tau_{BE}(\nu/\nu_{BE})^{-3}. \]
The best fit results are shown in Figure 2 and in Table 1.
The 68\% ($\sim 1\sigma$) confidence ellipses are also shown in Figure 3.
\begin{figure}
  \begin{center}
    \FigureFile(50mm,20mm){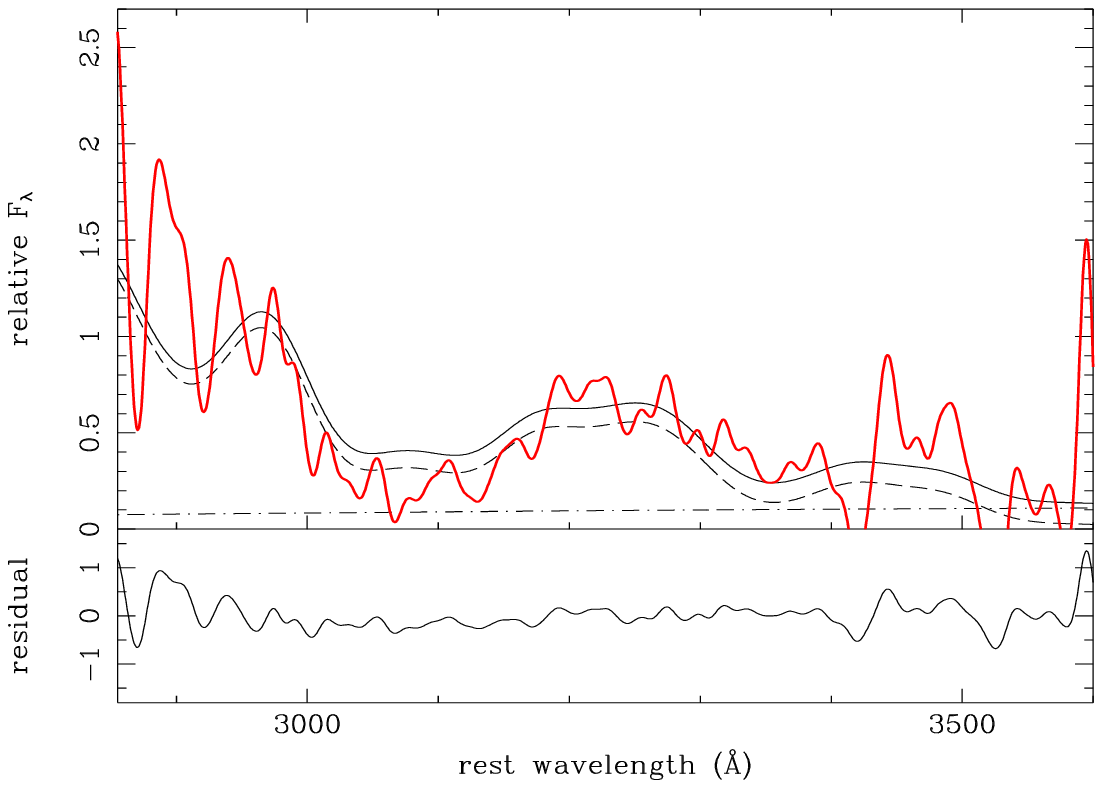}
  \end{center}
  \caption{The best fit result. The SDSS 1044-0125 spectrum with the power-law continuum subtracted is shown in a red line.
The Fe \emissiontype{II} model template smoothed with 6000 km/s FWHM and a Balmer 
continuum are shown with a dashed line and a dot-dashed line, respectively.
The Fe \emissiontype{II} plus a Balmer continuum is shown in a black solid line.}
\label{fig2}
\end{figure}
\begin{figure}
  \begin{center}
    \FigureFile(50mm,20mm){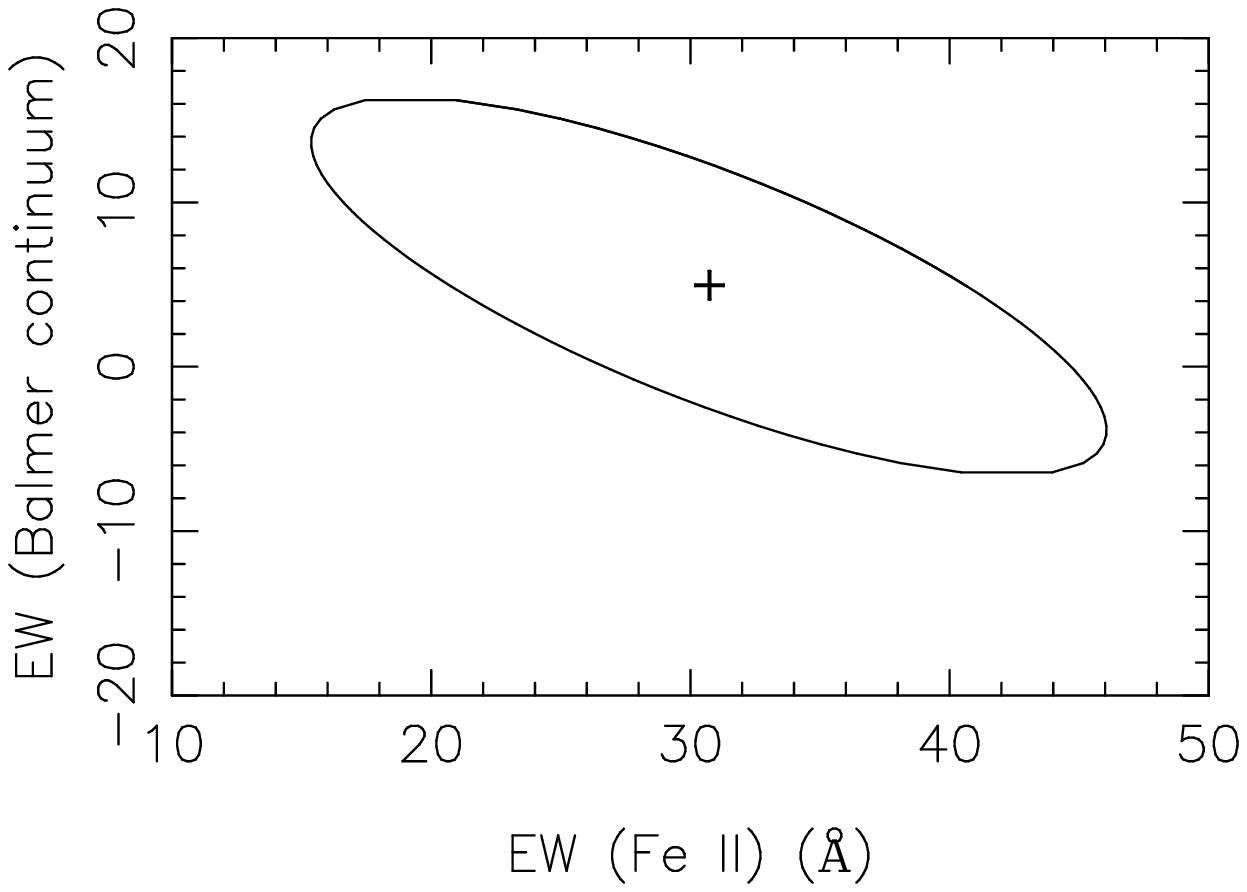}
  \end{center}
  \caption{The 68\% confidence ellipse for the fit.
The ordinate is rest frame EW of the Balmer continuum between 1000 \AA~ and 3650 \AA~ which is calculated from the scaling parameter $b$, the abscissa is the EW of the Fe II emission between 3000 \AA~ and 3500 \AA,  which is calculated from the scaling parameter $a$.
The crosses indicate the best fit value of EW.}
\label{fig2}
\end{figure}
\begin{table}
  \caption{The Result of Fit}\label{table1}
  \begin{center}
    \begin{tabular}{ccc}
      \hline
      $\alpha$ & $\chi^2$ (d.o.f.)& Fe \emissiontype{II}$^{a}$(rest EW (\AA)) \\
      \hline
      0.54 & 210 (52) & 31$\pm15$  \\
      \hline
      \multicolumn{3}{l}{$^{a}$ Fe \emissiontype{II} is measured between 3000 and 3500 \AA.} \\
    \end{tabular}
  \end{center}
\end{table}
\par
Combined with the power-law continuum flux at 3250 \AA, 
the rest equivalent width (EW) of Fe \emissiontype{II} (3000 - 3500 \AA) is $\sim$ 30 \AA.
There is a possibility of contamination of O \emissiontype{III} fluorescence line 
$\lambda3133$ and He \emissiontype{II} $\lambda3203$.
If the significant emission of O \emissiontype{III} $\lambda3133$ existed,
a 3200 \AA~ bump of the QSO would start $\sim 50$ \AA~ shorter wavelength than
the spectra of the SDSS 1044-0125.
We judged no significant contamination of O \emissiontype{III} fluorescence line.
We could not estimate of He \emissiontype{II} $\lambda3203$.
The residual at 3200 \AA~ may be He II (Fig.2).
However, a dip at 3300 \AA~ is not reproduced by only He \emissiontype{II} $\lambda3203$,
therefore we think that most of a bump between 3100 and 3400 \AA~ is Fe \emissiontype{II} emission.
We also measured Fe \emissiontype{II} strength of LBQS composite
spectrum 
and the composite quasar spectrum from the Sloan Digital Sky Survey (SDSS) (Vanden Berk et al. 2001) with the same manner as for SDSS 1044-0125.
The power-law continua were decided at the minimum between 1450 and 1470 \AA~and the minimum between 2200 and 2250 \AA.
The indexes of power-law continuum for LBQS and SDSS composites are 0.48 and 0.56, respectively.
We have done $\chi^2$ fitting between 2855 and 3600 \AA~ using the same Fe \emissiontype{II} template and the Balmer continuum using for SDSS 1044-0125.
The derived Fe \emissiontype{II} (3000 - 3500 \AA) EW of LBQS composite and 
that of SDSS composite are 32 \AA~ and 34 \AA, respectively.
This value is same as that of SDSS 1044-0125, so the strength of 
Fe \emissiontype{II} (3000 - 3500 \AA)
in SDSS 1044-0125 is similar to that of low redshift QSOs.
\par

\section{Discussion}
We could not observe the Mg \emissiontype{II} $\lambda2798$ emission line, however,
N \emissiontype{V}, C \emissiontype{IV}, and C \emissiontype{III}] emission-line strength 
in SDSS 1044-0125 are similar to low redshift QSOs (Goodrich et al. 2001; Maiolino et al. 2001). 
This fact suggests the strength of Mg \emissiontype{II} of SDSS 1044-0125 will be also 
similar to that of low redshift QSOs.
We measured the intensity of the Mg \emissiontype{II} emission line in LBQS composite and 
SDSS composite.
The $I$(Fe \emissiontype{II} (3000-3500 \AA)/$I$(Mg \emissiontype{II}) in LBQS composite 
and SDSS composite are 0.69 and 0.70, respectively.
Although we assume the strength of Mg \emissiontype{II} of SDSS 1044-0125 will be 
similar to that of low redshift QSOs and there is 50\% error in the measurement of Fe
\emissiontype{II} of SDSS 1044-0125,
the $I$(Fe \emissiontype{II} (3000-3500 \AA))/$I$(Mg \emissiontype{II}) in SDSS 1044-0125
will be similar at least or larger than the result 
(0.15-0.4) from solar abundance Broad Line Region model calculation (Wills et al. 1985).
Since the redshift of 5.74 corresponds to only 1.0 Gyr past from the Big Bang,
iron abundance of the QSO is predicted to be solar value assuming high star formation 
efficiency (Hamann \& Ferland 1993; Yoshii et al. 1998) which can reproduce 
$I$(N \emissiontype{V}) / $I$(He \emissiontype{II}) in high redshift QSOs (Hamann and Ferland 1993) and iron overabundance in low redshift QSOs.
Fe \emissiontype{II} emission in SDSS 1044-0125 is factor$\sim2$ stronger than the prediction by 
the chemical enrichment model of QSOs.
These chemical enrichment models assume the typical lifetime of the progenitor of SNe Ia is more than 1 Gyr, therefore iron enrichment relative to $\alpha$-element delays.
The 0.5-3 Gyr lifetime of the progenitor of SNe Ia naturally explains the [O/Fe] - [Fe/H] relation in the solar neighborhood (Yoshii et al. 1996; 
Kobayashi et al. 1998), however, the lifetime of the progenitor of SNe Ia is 
not theoretically constrained so much because the lifetime of the companion in case of a white dwarf model accreting from the matter from the companion or intrinsic separation in case of merging white dwarfs model (Iben, \& Tutukov 1984; Tutukov \& Yungelson 1994) is unknown.
The merging white dwarf model predicts shorter lifetime (0.3 - 0.4 Gyr) of the progenitor (Tutukov \& Yungelson 1994).
All observed high redshift ($z > 3$) QSOs which are all luminous one and must be massive galaxy have similar $I$ (Fe \emissiontype{II})/$I$ (Mg \emissiontype{II}) ratio to low redshift QSOs and no systematic decrease from them (Kawara et al. 1996; Murayama et al. 1999; Thompson et al. 1999; Dietrich and Hamann 2001; Dietrich et al. 2002).
This line ratio suggests iron overabundance even if the age of universe $\sim 1$ Gyr although there might be hidden physics which control Fe \emissiontype{II} emission line strength to be constant otherwise abundance change.
The progenitor of SNe Ia in the nucleus of QSOs where gravitational potential is deeper and star density is higher may be different from that in the solar neighborhood (Thompson et al. 1999). \par
Another interpretation of strong Fe \emissiontype{II} emission at $z=5.74$ is different cosmological parameters from those we adopted.
The relation of age and redshift depends on the cosmological parameters, the Hubble constant $H_{\rm 0}$, and the matter density $\Omega_{\rm M}$.
In a flat universe the age $t \propto H_{\rm 0}^{-1} \Omega_{\rm M}^{-1/2}$ at 
high redshift regime ($z > 3$).
If we assume $H_{\rm 0}$ to be 50 km sec$^{-1}$ Mpc$^{-1}$ , the cosmological age at $z=5.74$ will be 1.4 Gyr which is consistent with the model predicts iron overabundance (Yoshii et al. 1998).
The $H_{\rm 0}$ of 50, however, is too small than that derived from such as
 Cephid distance, Tully-Fisher relation, fundamental plane of elliptical galaxies,
surface brightness fluctuations and SNe Ia (Mould et al. 2000).
The estimate of the matter density $\Omega_{\rm M}$ , on the other hand, has rather large uncertainty. 
The measurement of inhomogeneity of the cosmic background radiation (de Bernardis et al. 2000; Melchiorri et al. 2000) and observations of high redshift SNe Ia (Schmidt et al. 1998; Permutter et al. 1999) indicate that a flat universe and $0.1 \lesssim \Omega_{\rm M} \lesssim 0.4$ (de Bernardis et al. 2000).
If we adopt $\Omega_{\rm M}$ of 0.2 instead of 0.3,
the cosmological age at $z=5.74$ will extends to 1.3 Gyr.
The large $I$(Fe \emissiontype{II})/$I$(Mg \emissiontype{II}) may suggest 
smaller $\Omega _{\rm M}$.
\section{Conclusion}
We obtained near-infrared spectroscopy of the $z=5.74$ QSO, SDSS 1044-0125
with the Infrared Camera and Spectrograph of the Subaru telescope.
We found the strength of Fe \emissiontype{II} emission to be similar to low redshift QSOs.
This is the highest redshift detection of iron.
Its equivalent width of
Fe \emissiontype{II} (3000-3500 \AA) is $\sim 30$ \AA~ which is similar to those of low redshift QSOs.
We estimated $I$(Fe \emissiontype{II} 3000-3500 \AA)/$I$ (Mg \emissiontype{II}) $\sim 0.7$.
Because the redshift of 5.74 corresponds to age of 1.0 Gyr in case 
of the current $\Lambda$-dominated cosmological parameter, this ratio is larger than prediction of chemical 
enrichment models that assume the life time of the progenitor of SN Ia is 
longer than 1 Gyr.
There is no systematic decrease of Fe \emissiontype{II} emission in observed high 
redshift QSOs ($z > 3$) from low redshift ones.
This suggests short lived SNe Ia progenitors may be more common in QSO nucleus than in the solar neighborhood.
It is necessary to search whether there is no hidden physics of Fe \emissiontype{II} emission mechanism in order to show definitely the possibility of short-lived SNe Ia progenitors. 
Another interpretation of high $I$(Fe \emissiontype{II})/$I$ (Mg \emissiontype{II}) 
at $z=5.74$ is smaller $\Omega_{\rm M}$ than we adopt.
The cosmological age at $z=5.74$ may be longer than 1 Gyr.
\par
We are grateful to the IRCS instrument team, especially H. Terada and 
N. Kobayashi for their help and valuable comments during our observations.
We also thank B. Wills for kindly providing an Fe \emissiontype{II} model template, her valuable comments, and improving the English expression.
Discussions with M. Yoshida, T. Yamada, and N. Arimoto have improved this 
paper.
The data reduction was done using the facilities of the Astronomical Data 
Analysis Center, National Astronomical Observatory of Japan, which is an inter-university 
research institute operated by the Ministry of Education, Culture, Sports, Science and Technology.
This work was financially supported in part by Grant-in-Aids
for the Scientific Research (No. 13740122) or the
Japanese Ministry or Education, Culture, Sports, Science and Technology.

\end{document}